\documentclass[bibliography=totoc]{scrartcl}
\usepackage[T1]{fontenc}
\usepackage[utf8]{inputenc}
\usepackage[]{amsmath,amssymb, amsthm}
\usepackage[round]{natbib}
\usepackage{graphicx}
\usepackage{subfigure}
\usepackage{bbm}
\usepackage{paralist}
\usepackage{blkarray}
\usepackage{multirow}
\usepackage{color}
\usepackage[toc,page]{appendix}
\usepackage{url}
\usepackage{epstopdf}
\usepackage{makecell}
\usepackage{lscape}
\usepackage[table,usenames,dvipsnames]{xcolor}
\usepackage{ltablex}
\usepackage{booktabs}
\usepackage{anysize}
\usepackage[pass]{geometry}

\begin{document}

	\author{Svenja Fischer 
		\thanks{SPATE research unit, Ruhr-Universit\"at Bochum,
			D-44801 Bochum, Germany,
			\texttt{svenja.fischer@rub.de}}}
	
	\date{}
	\title{Comparison of annual maximum series and flood-type-differentiated mixture models of partial duration series}
	\maketitle
	\bibliographystyle{plainnat}
	
	\setlength{\parindent}{0pt}
	\allowdisplaybreaks

\theoremstyle{definition}

\newcommand{\cov}{\text{Cov} } 
\newcommand{\var}{\text{Var} } 
\newcommand{\E}{\mathbb{E} } 
\newcommand{\1}{\mathbbm{1}}
\newcommand{\rdots}{\hspace{.2ex}\raisebox{1ex}{\rotatebox{-12}{$\ddots$}}}
\newcommand{\med}{\operatorname{med}}
\newcommand{\argmin}{\mathop{\mathrm{argmin}}}

\begin{abstract}
	\noindent
	The use of the annual maximum series for flood frequency analyses limits the considered information to one event per year and one sample that is assumed to be homogeneous. However, flood may have different generating processes, such as snowmelt, heavy rainfall or long-duration rainfall, which makes the assumption of homogeneity questionable. Flood types together with statistical flood-type-specific mixture models offer the possibility to consider the different flood-generating processes separately and therefore obtain homogeneous sub-samples. The combination of flood types in a mixture model then gives classical flood quantiles for given return periods. This higher flexibility comes to the cost of more distribution parameters, which may lead to a higher uncertainty in the estimation. This study compares the classical flood frequency models such as the annual maximum series with the type-specific mixture model for different scenarios relevant for design flood estimation in terms of Bias and variance. Thee results show that despite the higher number of parameters, the mixture model is preferable compared to the classical models, if a high number of flood events per year occurs and/or the flood types differ significantly in their distribution parameters.

\end{abstract}

\section{Introduction}

For many practical applications and theoretical research in flood statistics, the annual maximum series (AMS) is the most widely applied approach. The AMS is also among the recommended procedures to perform flood frequency analyses for many countries like the US \citep{Bulletin17C}, the UK \citep{FloodEstHandbook} or Germany \citep{DWA.2012}. Due to its simple application and few assumptions and requirements on the data, it is particularly often used by authorities, consulting engineers and other stake holders. For the AMS approach, the largest flood peak in a year is taken (the so-called block maximum) and a distribution is fitted to the sample of all these maxima in all years of the observation period. The obtained flood quantiles for a given return period serve as design flood for the construction of, e.g., dams or flood protection systems. The range of possible distributions that can be fitted is large, e.g., there are six different distributions recommended for Germany in \citep{DWA.2012}, but mainly two distribution functions are used: the log-PearsonIII distribution, which is frequently used in the US, and the Generalized Extreme Value Distribution (GEV), which is among the most-common distributions in Europe. The latter also has statistical justification since it is the limiting distribution of block maxima (Fisher-Tippett Theorem). The parameters of these distributions are most frequently estimated with either Method of Moments, Maximum Likelihood or L-Moment estimators, the latter being most frequently used in hydrology due to their efficiency and robustness \citep{Fischer.2016}. 
However, there is rising concern if the use of the AMS is sufficient to characterise the full spectrum of floods correctly. On the one hand, the consideration of only one flood event per year limits the information used for statistical analyses. Therefore, several researchers in hydrology recommend to use the Peak-over-Threshold (POT) series instead, where each flood peak over a given threshold is considered \citep{Madsen.1997}. The challenge in the application of this approach lies in the determination of the threshold, which can be done either statistically or hydrologically \citep{Fischer.2016}. Moreover, it can introduce more uncertainty than the AMS approach for certain scenarios \citep{Madsen.1997}. Yet, it has found its way into some of the guidelines for flood statistics \citep{Bulletin17C, FloodEstHandbook}. The other concern that is raised in the context of the AMS and that also cannot be addressed by the POT approach is the question of homogeneity. A fundamental assumption of both approaches described above is the homogeneity of the sample, meaning that the flood peaks arise from a set of common forces. This assumption goes back to the early works on flood statistics by \cite{Gumbel.1941}. However, it is highly questionable if flood events, that typically arise from different meteorological and catchment conditions such as heavy rainfall or snowmelt, can be seen as homogeneous. To overcome this problem, many works propose the splitting of the sample into floods of different origin, e.g., summer and winter floods, and combine them in a two- (or more) component mixture model \citep{Waylen.1982,	Allamano.2011,Rossi.1984,Cunderlik.2004}. The splitting guarantees that floods with similar flood-generating processes are considered jointly such that homogeneity of the samples can be obtained. A step further go the works of \cite{Fischer.2018} and \cite{Fischer.2019}, who apply a flood typology that directly considers the meteorological circumstances as well as the flood hydrograph shape and combine the flood-type specific samples in a mixture model, the type-based mixture model of partial duration series (TMPS).
The results demonstrate that a consideration of flood types improves the understanding of flood generation and provide more detailed information on the design flood compared to the classical approaches based on AMS and POT.

But, when applying such complex models with distinctly more parameters, the question of uncertainty arises. This makes detailed investigations of the suitability of the proposed model and the uncertainty introduced by its application necessary. Typically, such simulations are performed for different combinations of parameters and sample sizes, to understand the impact of the sample moments as well as the sample uncertainty. For example, \cite{Madsen.1997} compare the POT and the AMS model for different shape parameters of the GEV model, different sample sizes as well as different return periods. In this paper, we will build on the benchmark methodology by \cite{Madsen.1997} and extend the procedure to make it suitable for a comparison of AMS, POT and the TMPS model. The results emphasize that the more different the flood types, the less suitable are classical models AMS and POT. The error in estimation for AMS and POT increases with increasing sample size and decreasing return period. The paper is structured as follows: first, the basic methodology of all three models, TMPS, AMS and POT is introduced, as well as the simulation procedure. Second the results are discussed for different parameter scenarios and the results are compared. Finally, recommendations are given in the concluding section on when to use which model for flood frequency analyses.

\section{Methodology}
In this section, we will shortly introduce the basic statistical concepts of the two approaches that will be compared: the annual maximum series (AMS) and the type-specific mixture model of partial duration series (TMPS). Both models will be compared in Monte Carlo simulation studies. 

\subsection{Type-specific mixture model of partial duration series: TMPS}
The type-specific mixture model of partial duration series (TMPS) considers a sample of hydrograph peaks of different flood types. Such samples can, e.g., be derived by application of a flood event separation and a flood typology \citep{Fischer.2019}. The events and hence the peaks are treated as independent due to the separation. The whole sample of events then is split into sub-samples, one for each flood type. A statistical model that can describe the annual mixture distribution of these flood types was proposed by \citet{Fischer.2018}. We introduce the model in the following.

Consider that there exist $n_t$ flood types resulting in samples of peaks $X_{1;j},\ldots,X_{N_j;j}$, $j=1,\ldots, n_t$, where $N_j$ is the number of events of type $j$. For each flood type, a threshold $u_j$ is defined, for which each exceedance indicates a flood event. That is, if $X_{i;j}>u_j$, then $X_{i;j}$ is a flood peak. The distribution of all exceedances of $u_j$ for flood type $j$, $G_j(x;\theta_j,u_j)$, is modelled by the Generalized Pareto distribution (GPD) with parameter set $\theta_j$. The GPD is particularly suitable for modelling such POT series, since mathematically theory (Balkeema-de-Haan Theorem) implies the convergence of such series to the GPD. It was used before for modelling POT series in the hydrological context \citep{wang.1991,singh.1995,gharib.2017,Madsen.1997}.

The distribution function of the GPD with parameter set $\theta=(\kappa,\beta)$ and threshold $u$ is given by
\begin{align*}
	G(x;u)=\begin{cases}
		1-\left(1+\kappa\left(\frac{x-u}{\beta}\right)\right)^\frac{-1}{\kappa},&\kappa\neq 0\\
		1-\exp\left(-\frac{x-u}{\beta}\right),&\kappa=0
	\end{cases}
\end{align*}
with threshold $u$, scale-parameter $\beta>0$ and support $x\geq u$ for $\kappa\geq 0$ and $u\leq x\leq u-\beta/\kappa$ for $\kappa<0$. The special case $\kappa=0$ corresponds to the exponential distribution. Often, $\kappa$ is called the tail index, referring to the exceedances as tail of the distribution. The GPD has the property to be threshold-stable, meaning that a truncated GPD is still a GPD with same shape parameter as before.

Besides the distribution of the exceedances, for the TMPS also the probability of non-exceedance of the threshold is required. This can be derived by considering the distribution of all peaks $F_j$ of flood type $j$.

Combining these two distributions and all flood types $j=1,\ldots,n_t$ to obtain the distribution of the annual maxima, the TMPS model then is defined as

\begin{align}\label{TMPS}
	H(x)=\prod_{j=1}^{n_t}\left(G_j(x;\theta_j,u_j)(1-F_j(u_j))+F_j(u_j)\right).
\end{align}

\subsection{Annual maximum series and Peak-over-Threshold}

The annual maximum series (AMS) is defined as the sample that contains the maximum flood peak of each year. It is among the most commonly used methods for flood frequency analysis. This has several reasons. First, AMS are more easily available than higher resolution samples. The calculation of annual return periods, which are required for design purposes, is straightforward and most models that are used to describe the AMS are therefore comparably simple with only few parameters. Moreover, when only taking one event per year, the independence assumption can assumed to be fulfilled (as long as not two peaks of consecutive days in two different years are chosen). However, the AMS does not consider the possibility of different origins of floods and therefore the inhomogeneity of the sample. When having in mind the sample of flood events of different types from above, the AMS is simply the maximum applied to each year. Due to this definition, the GEV has a mathematical justification (Fisher-Tippett-Theorem), since it is the limiting distribution of block maxima (under specific assumptions).

The distribution function of the GEV is given by 
$$F(x)=\exp\left(-\left(1+\xi\frac{x-\mu}{\sigma}\right)^{-\frac{1}{\xi}}\right)$$ 	
for $1+\xi(x-\mu)/\sigma>0$, where $\xi \in \mathbb{R}$ is the shape parameter, $\sigma>0$ is the scale parameter and $\mu\in \mathbb{R}$ is the location parameter. The special case $\xi=0$ is the Gumbel distribution.

Moreover, the GEV is obtained when considering the GPD as distribution of the POT series and combining it with the Poisson distribution to describe the occurrence of the number of events per year such that the annual distribution $F_a$ is derived, the so-called Poisson-Pareto model \citep{Stedinger.1993}, here simply referred to as POT approach:

\begin{align*}
	{{F}_{a}}(x)=\sum\limits_{k=1}^{\infty }{\frac{{{\lambda }^{k}}}{k!}}{{e}^{-\lambda }}{{\left( 1-\left( 1+\kappa \frac{x-{u}}{\beta } \right)-\frac{1}{\kappa } \right)}^{k}}={{e}^{-\lambda }}\exp \left( \lambda \left( 1-{{\left( 1+\kappa \frac{x-{u}}{\beta } \right)}^{-\frac{1}{\kappa }}} \right) \right)  \\
	=\exp \left( -\lambda {{\left( 1+\kappa \frac{x-{u}}{\beta } \right)}^{-\frac{1}{\kappa }}} \right).
\end{align*}
with parameters $\xi=\kappa$, $\sigma=\beta\lambda^\kappa$ and $\mu =u-\beta (1-{{\lambda }^{\kappa }})/\kappa$.

Roughly spoken, the main difference between the AMS with GEV distribution and the POT with Poisson-Pareto model lies in the information contained in the data below the POT-threshold.

\subsection{Parameter estimation}

In this study, we will limit the number of variations in the simulation study by only comparing one parameter estimation, the $L$-moments. $L$-moments are among the most often applied parameter estimators in hydrological statistics. They have the general advantage of being more robust than, e.g., the Maximum-Likelihood estimation and at the same time being efficient also for small samples. They exist for distributions where the classical moments do not exist. $L$-moments origin from a linear combination (l(inear)-moments) of the probability weighted moments \citep{HoskingJ.R.M..1990}.

\vspace{0.3cm}

Let $X$ be a real-valued random variable with cumulative distribution function $F$ and quantile function $x(F)$. The $r$-th $L$-moment is defined as
\begin{equation}
	\lambda_r=\frac{1}{r}\sum_{k=0}^{r-1}(-1)^k\binom{r-1}{k}\mathbb{E}X_{(r-k:r)},
\end{equation}
where $X_{(i:n)}$ is the $i$-th value of the order statistics of a sample $X_1,\ldots,X_n$ drawn from $F$ and $r=1,2,\ldots$. For example, the first two $L$-moments are given by
\begin{align*}
	\lambda_1=\mathbb{E}X&=\int_0^1x(F)dF,\\
	\lambda_2=\frac{1}{2}\mathbb{E}\left(X_{(2:2)}-X_{(1:2)}\right)&=\int_0^1x(F)(2F-1)dF.
\end{align*}

\citet{Hosking.1986} has shown that $\lambda_r$ exists for $r=1,2,\ldots$ if and only if $\mathbb{E}\vert X\vert$ exists.

\vspace{0.3cm}

The sample $L$-moment then is given by
\begin{align*}
	l_r&=\frac{1}{\binom{n}{r}}\sum_{1\leq i_1<\ldots < i_r\leq n}r^{-1}\sum_{k=0}^{r-1}(-1)^k \binom{r-1}{k}x_{(i_{r-k}:n)}\\
	&=\sum_{k=0}^{r-1}(-1)^{r-1-k}\binom{r-1}{k}\binom{r-1+k}{k}\frac{1}{n}\frac{1}{\binom{n-1}{k}}\sum_{j=k+1}^n\binom{j-1}{k}x_{(j:n)}.
\end{align*}

$l_r$ is an unbiased estimator for $\lambda_r$. Besides the simple $L$-moments, often the so-called $L$-moment ratios are used

\begin{align}\label{SecParEst}
	\tau_r=\lambda_r/\lambda_2,
\end{align}
respectively their sample estimates with the special cases of $\tau_3$, the $L$-skewness, and $\tau_4$, the $L$-Kurtosis.

\vspace{0.3cm}

For right-skewed samples, referring to a shape parameter larger than 0.2, which is often the case for samples of flood peaks, \citet{Madsen.1997} show that for the AMS as well as the POT series the $L$-moment estimates are preferred concerning Root Mean Squared Error (RMSE) for small and moderate sample sizes as well as return periods. Similar results have been reported by \citet{Hosking.1986} and \citet{Hosking.1985} for POT and AMS series. The Method of Moments estimates, instead, often led to a large negative Bias for a shape parameter larger than 0.3, while the maximum-likelihood estimator does not converge for small sample sizes and small shape parameter \citep{Madsen.1997}. Based on these results, we decided to only apply the $L$-moment estimators for the distribution parameters to reduce the complexity of the simulation study. We will consider small and large sample sizes and return periods and therefore require a reliable estimate for all these ranges.

\vspace{0.3cm}

For the parameters of the GEV distribution, the $L$-moment estimators are \citep{HoskingJ.R.M..1990}

\begin{align*}
	\hat{\xi }&=-7.8590\cdot f-2.9554\cdot {{f}^{2}}\\
	\text{with  }	f&=\frac{2{{l}_{2}}}{{{l}_{3}}+3{{l}_{2}}}-\frac{\ln 2}{\ln 3}\\
	\hat{\sigma }&=\frac{-\hat{\xi }\cdot {{l}_{2}}}{\Gamma \left( 1-\hat{\xi } \right)\left( 1-{{2}^{\hat{\xi }}} \right)}\\
	\hat{\mu }&={{l}_{1}}-\hat{\sigma}\left[ \Gamma \left( 1-\hat{\xi } \right)-1 \right]/\hat{\xi }.
\end{align*}

For the estimation of the parameters of the GPD distribution, two cases have to be distinguished: if the parameter of location is known or unknown.

In this work, the location parameter, i.e. the threshold $u$, will always assumed to be known. Hence, the following two estimators are given \citep{pandey.2001}.

\begin{align*}
	\hat{\kappa}&=2-\frac{l_1-u}{l_2}\\
	\hat{\beta}&=(1-\hat{\kappa})(l_1-u).
\end{align*}

\subsection{Simulation Algorithm}

The aim of this study is to compare AMS and TMPS models concerning Bias and RMSE. In Monte Carlo simulations, different return periods and sample sizes are compared that lead to recommendations on when to use which model. However, it is not straightforward to design such Monte Carlo simulations. Already the comparison of POT and AMS led to several problems that were addressed by \citet{Madsen.1997}: the simulation of a POT sample from a given distribution such as GPD does not provide sufficient information about the AMS and vice versa. When simulating a POT sample, the probability that one event of the AMS is below the POT-Threshold is $\exp(-\lambda)$ when $\lambda$ is the mean annual number of events above the threshold. Similarly, the simulation of an AMS sample will exclude several POT events, especially if $\lambda>1$. To overcome this problem, they made use of the threshold-stability of the GPD.

Here, we will use a similar technique to design the Monte Carlo simulation. However, for the TMPS model an additional component has to be considered compared to the POT model: it makes use of the non-exceedance probability of the threshold $u_j$. This probability cannot be derived from a POT-sample and hence it is impossible to know the true value when performing a Monte Carlo simulation. Therefore, we decided to fix this probability for each flood type. Of course, this reduces the complexity of the TMPS model and hence a comparison with AMS may not be fully fair, but it is the only way to design such an experiment. Moreover, it only has small impact on the TMPS results, which are mainly influenced by the GPD distribution \citep{Fischer.2018}. We denote this non-exceedance probability with $p_{0;j}$ for each flood type $j=1,\ldots,n_t$. Though $p_0$ is somehow related to the threshold $u$, it cannot be estimated from $u$, since $p_0$ requires more information, also from the values below the threshold.

Additionally, we have to define the average number of events per flood type and year. This number is denoted by $\lambda_j$. This variable is not required for the TMPS model but to derive the AMS model from the sample, as shown later on.

With these assumptions, first a sample of size $n$ years is drawn from the GPD distribution for each flood type. For this purpose, $n\cdot\lambda_j$ values $X_{1;j},\ldots,X_{n\lambda_j;j}$ are generated from the GPD using the parameter set $\theta_j=(\kappa_j, \beta_j)$ and the threshold $u_j$. 

The true annual distribution of the joint sample $X_{1;1},\ldots,X_{n\lambda_1;1},\ldots,X_{1;n_t},\ldots,X_{n\lambda_{n_t};n_t}$ simply is the TMPS model with parameters $\kappa_1,\ldots,\kappa_{n_t},\beta_1,\ldots,\beta_{n_t},u_1,\ldots,u_{n_t}$ and $F_j(u_j)=p_{0;j}$.

Following \citet{Madsen.1997}, the corresponding AMS then can be derived by taking the maximum of each year and adding the threshold parameter $u$. However, since we consider $n_t$ different flood types, this has to be extended. In our case, the AMS of year $i$, $Y_i$, is defined as the maximum of the maximum value of each flood type per year plus type-specific threshold $u_j$:

$$Y_i=\max\limits_{1\leq j\leq n_t}\left(\max\limits_{(i-1)\lambda_j+1\leq k \leq i\lambda_j}X_{k;j}+u_j\right), i=1,\ldots,n.$$

In a next step, the three models, TMPS, POT and AMS, are fitted to the generated samples respectively. For the TMPS, the GPD distribution is fitted for each type separately using the $L$-moments. For the AMS, the GEV is fitted to the AMS sample using the L-moments. We also compare these two scenarios with the POT case as described in \citep{Madsen.1997}, where a GPD distribution is fitted to the POT sample. Then, the respective quantiles for a given return period $T$ are estimated. This procedure is repeated $N$ times and normalized Bias and RMSE are derived

\begin{align*}
	Bias(T)_i&=\frac{q_T-\hat{q}_T}{q_T}, i=1,\ldots,N\\
	RMSE_T&=\sqrt{\frac{1}{N}\sum_{i=1}^N(\frac{q_T-\hat{q}_T}{q_T})^2},
\end{align*}

where $q_T$ is the true quantile and $\hat{q}_T$ is the estimated quantile either from the TMPS or the AMS model for return period $T$.

\section{Results}
For the simulation study, different settings were considered. For each setting, 1000 runs were performed. A higher number of runs would be computationally too expansive due to the optimisation procedure used to derive the TMPS quantiles. Return periods $T$ were varied between $T=2,5,10,20,25,50,100,200,500$ years to consider the full spectrum of small and extreme quantiles and therefore possible heavy tails in the data. Sample sizes $n$ were varied between $n=30,50,100,200,500$ years and the case where asymptotics should be valid, $n=10000$.
We considered five different flood types, similar to the ones proposed by \citet{Fischer.2019}, to have a reasonable variation between the flood events. The probability of exceedance of the threshold were chosen as $p_0;1=0.24$, $p_{0;2}=0.20$, $p_{0;3}=0.17$, $p_{0:4}=0.26$ and $p_{0;5}=0.14$. These values have been obtained by averaging all probabilities of exceedances of the respective flood types for a large sample of gauges in Bavaria, Southern Germany \citep{Fischer.2021}. The variability in the probability can be considered in the analysis of the results.

All remaining parameters were varied in different scenarios.

\begin{enumerate}
	\item Scenario: Variation of shape, one event per year\\
	The number of events per year was chosen as $\lambda_j=1$ for all $j=1,\ldots,5$, the scale parameter was chosen as $\beta_j=5$ for all $j=1,\ldots,5$ and threshold parameter was set to $u_j=10$, $j=1,\ldots,5$. The shape parameter was first set to $\kappa_j=0.2$, but the number of flood types with $\kappa_j=0.6$, corresponding to a heavy tail, was increased successively from $n_E=0$ to $n_E=5$ in different simulations.
		\item Scenario: Variation of shape, two events per year\\
	The number of events per year was chosen as $\lambda_j=2$ for all $j=1,\ldots,5$, the scale parameter was chosen as $\beta_j=5$ for all $j=1,\ldots,5$ and threshold parameter was set to $u_j=10$, $j=1,\ldots,5$. The shape parameter was first set to $\kappa_j=0.2$, but the number of flood types with $\kappa_j=0.6$, corresponding to a heavy tail, was increased successively from $n_E=0$ to $n_E=5$ in different simulations.
		\item Scenario: Variation of scale\\
	The number of events per year was chosen as $\lambda_j=2$ for all $j=1,\ldots,5$, the shape parameter was chosen as $\kappa_j=0.2$ for all $j=1,\ldots,5$ and threshold parameter was set to $u_j=10$, $j=1,\ldots,5$. The scale parameter was first set to $\beta_j=5$, but the number of flood types with $\beta_j=20$, corresponding to a high variation, was increased successively from $n_E=0$ to $n_E=5$ in different simulations.
		\item Scenario: Variation of location\\
	The number of events per year was chosen as $\lambda_j=2$ for all $j=1,\ldots,5$, the shape parameter was chosen as $\kappa_j=0.2$ for all $j=1,\ldots,5$ and scale parameter was set to $\beta_j=5$, $j=1,\ldots,5$. The threshold parameter was first set to $u_j=10$, but the number of flood types with $u_j=50$, corresponding to a high mean value, was increased successively from $n_E=0$ to $n_E=5$ in different simulations.
			\item Scenario: Variation of shape and scale\\
	The number of events per year was chosen as $\lambda_j=2$ for all $j=1,\ldots,5$, the threshold parameter was set to $u_j=10$, $j=1,\ldots,5$. Both, the shape parameter and scale parameter were varied for an increasing number of flood types. The shape parameter was first set to $\kappa_j=0.2$, but the number of flood types with $\kappa_j=0.6$, corresponding to a heavy tail, was increased from 0 to 5. The scale parameter was first set to $\beta_j=5$, but the number of flood types with $\beta_j=20$, corresponding to a high variation, was increased successively from $n_E=0$ to $n_E=5$ in different simulations.
\end{enumerate}

\begin{figure}
\centering
	\includegraphics[width=\textwidth]{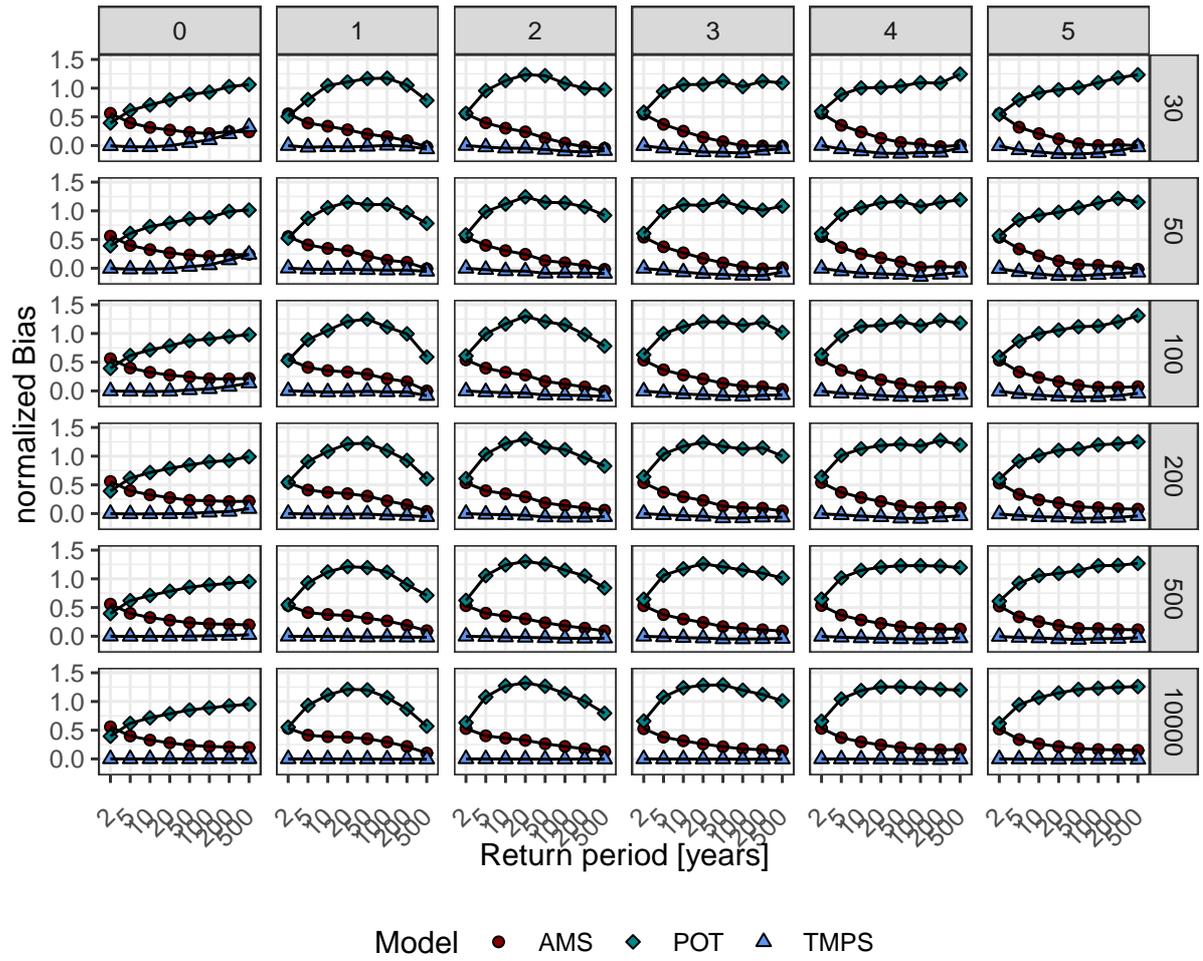}
		\caption{Mean normalized Bias for different return periods stratified by sample size $n$  (rows) and number of flood types with extreme parameters $n_E$ (columns) with AMS, POT and TMPS models in 1. Scenario.}
	\label{Fig:Scenario1}
\end{figure}

\begin{figure}
\centering
	\includegraphics[width=\textwidth]{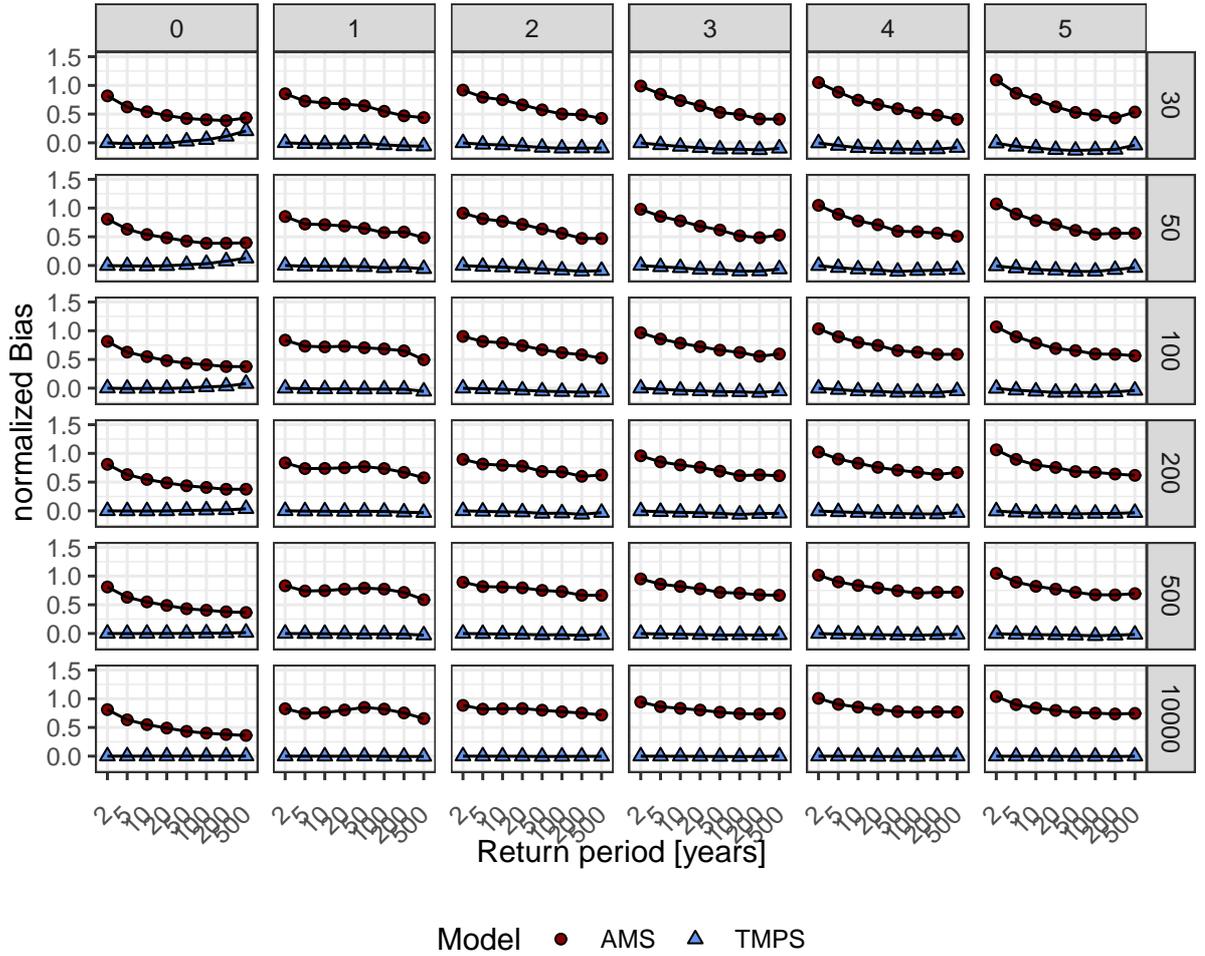}
		\caption{Mean normalized Bias for different return periods stratified by sample size $n$  (rows) and number of flood types with extreme parameters $n_E$ (columns) with AMS, POT and TMPS models in 2. Scenario.}
	\label{Fig:Scenario2}
\end{figure}

\begin{figure}
	\centering
	\includegraphics[width=\textwidth]{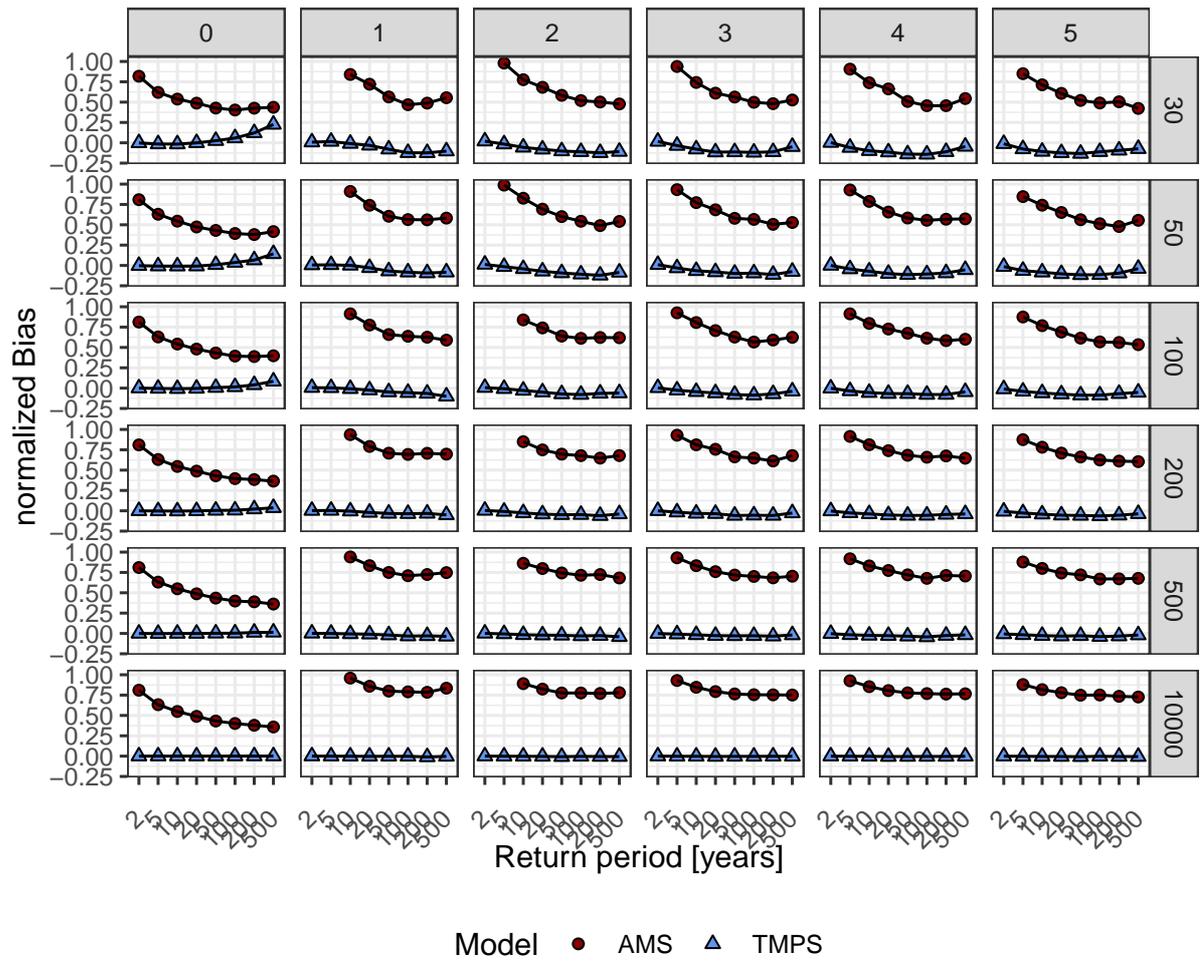}
	\caption{Mean normalized Bias for different return periods stratified by sample size $n$  (rows) and number of flood types with extreme parameters $n_E$ (columns) with AMS, POT and TMPS models in 5. Scenario.}
	\label{Fig:Scenario5}
\end{figure}

Results in terms of normalized Bias are given for different return periods and sample sizes in Figures \ref{Fig:Scenario1}-\ref{Fig:Scenario5} for Scenarios 1, 2 and 5, respectively Figures \ref{Fig:Scenario4} and \ref{Fig:Scenario5} in the Appendix for Scenarios 3 and 4. Mean normalized Bias as well as RMSE are given for selected Scenarios and conditions in Table \ref{Tab:RMSE} to obtain information on the variability of the estimation for 1000 repetitions. Due to the low performance results of the POT model in Scenario 1 (Figure \ref{Fig:Scenario1}), this model was no longer considered for the remaining scenarios.

\vspace{0.3cm}
For the first scenario, where the shape parameter was increased from $\kappa_j=0.2$ to $\kappa_j=0.6$ for a varying number of flood types $j$ and one flood event of each flood type was considered per year, the TMPS model resulted in a normalized Bias around zero (Figure \ref{Fig:Scenario1}). Only for small sample sizes of $n=30$ up to $n=100$, return periods greater than 50 years and no extreme shape parameter for any of the flood types, the Bias increased to a maximum of 0.4. Therefore, the TMPS is well capable for the estimation of the mixture distribution, even for small sample sizes and large return periods, especially if one or more flood types have a high shape parameter. The AMS model instead resulted in constantly high Bias of about 0.5 for small return periods of 2-5 years and all sample sizes. However, with increasing return periods the Bias reduced and for return periods of about 500 years the Bias of the AMS was comparable to that of the TMPS model. Similarly to the TMPS, the Bias for large return periods was smaller if one or more flood types had a large shape parameter. The POT is only comparable to the AMS for very small return periods up to 5 years. For larger return periods and despite of the sample size, it led to much higher Bias. The POT model performance suffered most from the increasing number of flood types with large shape parameters, where especially the Bias of the large return periods greater than 50 years increased. The very poor performance of the POT model in this case makes it unsuitable for the consideration of flood types. Therefore, it is no longer considered in the following scenarios.

\vspace{0.3cm}
For the second scenario, the number of flood events per year for each flood type was increased to $\lambda_j=2$, while the remaining setting stayed the same as in Scenario 1. In this case, the Bias of the TMPS model still stayed low around zero and was comparable to the first scenario (Figure \ref{Fig:Scenario2}). The Bias of the AMS model, however, increased significantly. Only for the case where none of the flood types had a large shape parameter, the Bias of the AMS and the TMPS model were comparable for large return periods and small sample sizes. The more flood types were assigned with large shape parameters and the larger the sample size, the higher became the Bias of the AMS model. For a number of five flood types with large shape parameter and a sample size of $n=100$, for example the Bias was never smaller than 0.5.

\vspace{0.3cm}

A distinct Bias of the TMPS model occurred in the third scenario, where again the number of flood events per flood type was set to $\lambda_j=2$ per year, but instead of the shape the parameter the scale parameter was altered from $\beta_j=5$ to $\beta_j=20$ (Figure \ref{Fig:Scenario3}). For this scenario, the Bias of the TMPS model increased for small sample sizes up to $n=100$ years, large return periods and for those scenarios, where (almost) all flood types had equal scale parameter, independent of if it was high or low. For all other cases, the Bias of the TMPS model was still around zero. Please note that the case where none of the flood types has increased parameter values is of course identical to the second scenario, when also none of the parameters was altered. The AMS model resulted again in higher Biases than the TMPS model, especially for small return periods. This Bias reduced with increasing return periods, independent of the sample size. The largest Bias occurred for those cases, where only one up to three flood types had large scale parameters.

Similar results can be observed for the fourth scenario, where instead of the scale parameter the threshold parameter was increased from $u_j=10$ to $u_j=50$ (Figure \ref{Fig:Scenario4}). While the TMPS has in general lower Bias in this Scenario, the AMS Model resulted in even higher Bias compared to Scenario 3.

\vspace{0.3cm}

Based on these results, for the final scenario the two parameters with largest impact on the results were varied at the same time: the shape and scale parameter. On the one hand, the shape parameter was increased from $\kappa_j=0.2$ to $\kappa_j=0.6$ for an increasing number of flood types and similarly the scale parameter was varied between $\beta_j=5$ and $\beta_j=20$ with again a number of two flood events per flood type and year. This way, high Bias can be expected for both AMS and TMPS since TMPS was most affected by the shape parameter and AMS by the scale parameter. Interestingly, except for small sample sizes of $n=30$ and $n=50$ years and high return period of more than 100 years together with small values for both parameters and all flood types, the TMPS was only slightly affected by the changes in parameters (Figure \ref{Fig:Scenario5}). The Bias was constantly low and around zero, even for small sample sizes. The AMS model instead was highly affected by this scenario constellation and the Bias was mostly above 0.5, no matter which return period or sample size was considered.
 
The results so far gave the Bias of the estimation with either the AMS or the TMPS model. However, it is also interest how high the variability of the estimation is to quantify the uncertainty. In Table \ref{Tab:RMSE}, this variability is given in terms of the RMSE. The RMSE for the TMPS model and small sample sizes is comparable to that of the AMS for almost all scenarios. Interestingly, despite of the higher number of parameters of the model, the RMSE of the TMPS decreases much faster for increasing sample sizes compared to the AMS. Yet, for both models the variability is high for the given special cases.

\begin{table}
	\caption{Normalized Bias and RMSE for selected scenarios and return period $T=100$ with the AMS or TMPS model.}
	\begin{tabular}{l c c c c c c c c}
		\hline
		no. extreme types & \multicolumn{4}{c}{AMS} & \multicolumn{4}{c}{TMPS}\\
		&\multicolumn{2}{c}{Bias} & \multicolumn{2}{c}{RMSE} &\multicolumn{2}{c}{Bias} & \multicolumn{2}{c}{RMSE}\\
		& n=30 & n=100 & n=30 & n=100 & n=30 & n=100 & n=30 & n=100\\
		\hline
		&\multicolumn{8}{c}{1. Scenario (varying $\kappa$ and $\lambda_j=1$)}\\
		$n_E=0$ & 0.210 & 0.211 & 0.293 & 0.182 & 0.099 &0.036 &0.250 & 0.137\\
		$n_E=2$ & 0.043 & 0.118 & 0.592 & 0.436 & -0.096 & -0.074 & 0.434 & 0.285\\
		&\multicolumn{8}{c}{2. Scenario (varying $\kappa$ and $\lambda_j=2$)}\\
		$n_E=0$ & 0.403 & 0.408 & 0.386 & 0.217 & 0.055 & 0.021 & 0.186 & 0.098\\
		$n_E=2$ & 0.501 & 0.617 & 0.861 & 0.598 & -0.095 & -0.061 &0.323 & 0.219\\
		& 	\multicolumn{8}{c}{3. Scenario (varying $\beta$ and $\lambda_j=2$)}\\
		$n_E=2$	&0.395 &0.393 & 0.491 & 0.292 & 0.024 & 0.006 & 0.252 & 0.130\\
		&\multicolumn{8}{c}{4. Scenario (varying $u$ and $\lambda_j=2$)}\\
		$n_E=2$ &0.645 & 0.664 & 0.206 & 0.132 & 0.017 & 0.009 & 0.118 & 0.068\\
		&\multicolumn{8}{c}{5. Scenario (varying $\kappa$, varying $\beta$ and $\lambda_j=2$)}\\
		$n_E=2$ &0.519 &0.612 & 0.893 & 0.630 & -0.110 & -0.079 & 0.411 & 0.241\\
		\hline
	\end{tabular}
\label{Tab:RMSE}
\end{table}

\section{Discussion}

The simulations demonstrated that there exist large differences in the estimation error between the AMS and the TMPS model which depend much on the scenario. The basis for the simulation was a mixed sample of different flood types with different parameter sets. The aim of this study was to find out if the AMS is suitable to model such mixed sample.

The results revealed that there are indeed scenarios, where the AMS delivered comparable results in terms of Bias to the TMPS model. This was especially the case if only one flood event per year and flood type occurred and for large return periods. For these cases, the value of the distribution parameter of the flood types was irrelevant for the estimation error. The AMS model is able to capture the distribution of the distribution of the dominant flood type if only one or two flood types have large shape parameters, since in this case the annual maximum will probably be of this flood type. The large shape parameter leads to extreme floods of this type which then will be also the annual maximum. However, if all flood types have similar parameters, there is no difference between the sample and it does not make a difference which flood type is represented in the AMS. Due to the simplicity of the AMS model compared to the TMPS model, we recommend the application of the AMS in this case, especially for the estimation of design floods of large return periods. The scenario of only one flood event per year and type may occur in arid regions and therefore the AMS model may be sufficient there.
However, the results revealed also that an increase of the number of flood events per year led to large differences between the AMS and the TMPS model and that the TMPS model resulted in much smaller Bias around zero. 
These differences increased for increasing shape parameters of the distribution of exceedances of the flood types. This behaviour of the AMS model implies that it cannot include the increased information that is obtained when having more than one flood event per year and flood type. Since the AMS approach only considers one event per year, it looses information: the larger the shape parameter and the more flood types are assigned with high shape parameters, the more large and extreme floods are generated. However, the annual maximum series can only consider one event per year and therefore many large events are ignored. This led to higher Bias of the AMS. This performance of the AMS is worsened, if not the shape but the scale and location parameters are increased. These two parameters increase the variability and the general level of the flood peaks and the information loss by only considering one event per year is highest. The Bias of the TMPS model, instead, increased most when having increasing shape parameters. The uncertainty introduced by the increased shape parameters multiplies due to the many parameters that have to estimated for this model. Therefore, the TMPS model is very sensitive to high shapes. Nevertheless, the Bias was still below the Bias of the AMS.
The more parameter are increased and therefore the higher the large floods and especially the variability, the worse is the performance of the AMS.
On the other hand, the results also revealed that the AMS model had problems in modelling highly different flood types. The less equal the distributions of the flood types were, the higher was the Bias of the AMS model. Again, the selection of only one event per year and therefore only one flood type does not mirror the complex distribution of many flood types correctly.
Therefore, for regions with highly unequal distributions of flood types and where large parameters for one or two flood types can be expected, we cannot recommend the use of the AMS model since the Bias is too high. This leads to high underestimation (the Bias was positive), especially of extreme flood peaks and have serious impact on the estimation of design floods and flood protection. This is the case for many regions around the globe, e.g. in Central Europe where floods caused by heavy rainfall seem to become more extreme than other flood types. Changing climate can increase this problem.

\vspace{0.3cm}

Often, when applying more complex distributions such as the TMPS, it is argued that the smaller Bias is only due to over-parametrisation and comes at the cost of high variability of the estimation. However, the results showed that the variability of both models, AMS and TMPS is almost equal for most scenarios. On the contrary, for large sample sizes the TMPS model has smaller RMSE. The AMS model has constantly high uncertainty since it only considers the samples partially and the consideration of floods depends much on their respective occurrence in single years.

\vspace{0.3cm}

It has to be noticed that the proposed simulation scenarios only give a limited view of the variability for varying parameters due to the limited number of variations of parameters. However, because of the complex model of the TMPS with its many parameters, a more detailed analysis is not easily possible. The proposed scenarios mirror the most important variations and therefore provide information on almost all important cases. Still, the simulation is limited and, e.g., the assumption of fixed probabilities of non-exceedance of the threshold in the TMPS model may reduce the RMSE and the Bias of the TMPS model. A variation of these parameters, too, is not possible. The large differences between the Bias of AMS and TMPS cannot be explained by this limitation and therefore the results remain significant.

\section{Conclusions}

The type-based mixture model of partial duration series (TMPS) offers the possibility to consider flood types separately and to combine them in a mixture distribution. It is argued that this leads to homogeneous samples and increases the included information in the model compared to classical approaches such as the annual maximum series (AMS). However, this comes of the cost of more parameters and therefore a potential increased uncertainty in the estimation of design floods. This work aimed to compare both, the TMPS and the AMS model, in terms of Bias and RMSE for selected and common parameter scenarios. It was shown that the TMPS outperforms the AMS model for all scenarios where the parameters were large and therefore lead to increased flood peaks and several flood types were affected. The largest differences occurred for small and medium return periods up to 100 years. The sample size did not play a crucial role. 

\vspace{0.3cm}

Based on these results we recommend to use the AMS model only for regions where the different flood types have similarly small shape and scale parameters and where approximately only one per year and flood type occurs. A large number of flood events per year leads to Bias of the AMS since only one event per year is considered in this model. Moreover, if one or two flood types differ significantly from the remaining one in terms of their distribution parameters, the AMS cannot be recommended, too. In this case, the AMS cannot incorporate the inhomogeneity of the samples and the statistical assumptions of the model are violated. This leads to significant under-estimation of design floods, as was shown above. Especially when having in mind recent large floods caused by heavy rainfall events in Central Europe, which clearly show that the distribution of this flood type differs significantly, e.g., from snowmelt floods, the wide-spread use of the AMS for the estimation of design floods should be re-considered.

Though the scenarios did not consider all types of parameter variation, the large differences in the Bias of AMS compared to the TMPS model gave occasion to question the wise-spread application of the AMS model in flood frequency analysis.

\section{Acknowledgements}
The financial support of the Deutsche Forschungsgemeinschaft (FOR2416, Research Unit SPATE ”Space-Time Dynamics of Extreme Floods”) is gratefully acknowledged. The author also would like to thank Richard Vogel for his helpful suggestions and comments that led to this work.

\nocite{*}
\bibliography{References}

\begin{appendices}
	\begin{figure}
		\centering
		\includegraphics[width=\textwidth]{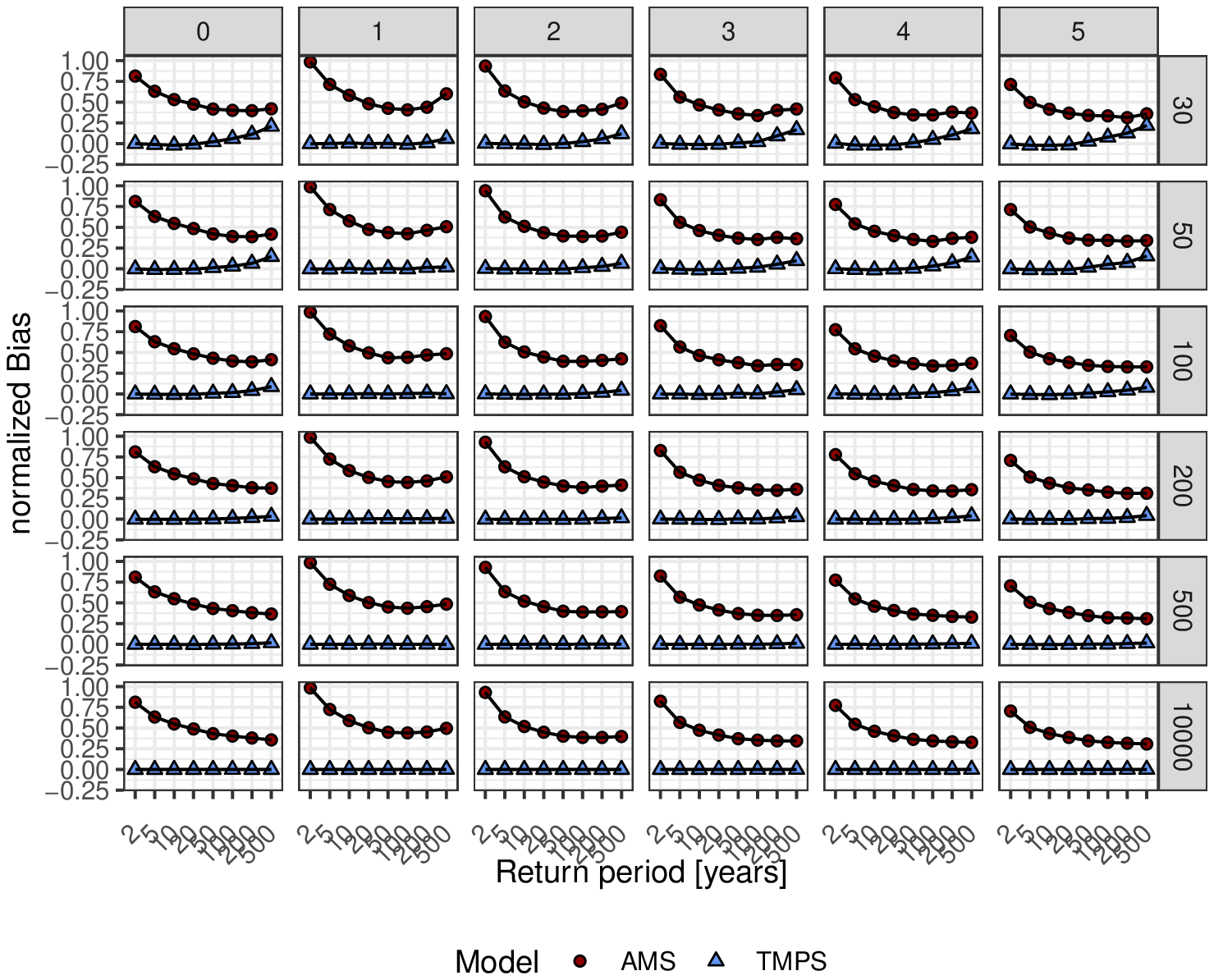}
		\caption{Mean normalized Bias for different return periods stratified by sample size $n$  (rows) and number of flood types with extreme parameters $n_E$ (columns) with AMS, POT and TMPS models in 3. Scenario.}
		\label{Fig:Scenario3}
	\end{figure}

	\begin{figure}
	\centering
	\includegraphics[width=\textwidth]{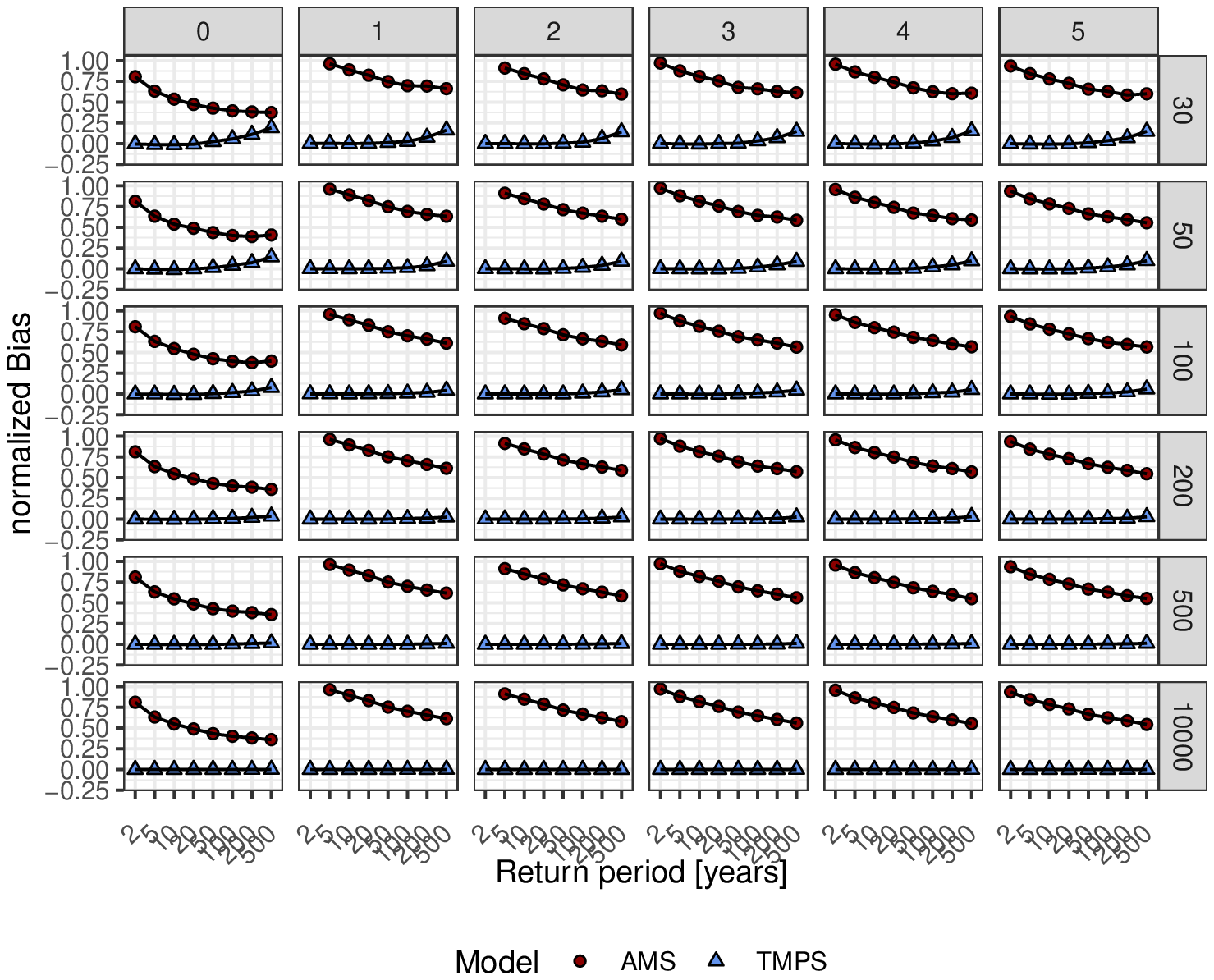}
	\caption{Mean normalized Bias for different return periods stratified by sample size $n$  (rows) and number of flood types with extreme parameters $n_E$ (columns) with AMS, POT and TMPS models in 4. Scenario.}
	\label{Fig:Scenario4}
\end{figure}

\end{appendices}

\end{document}